\documentclass[aps, prl, showpacs, showkeys, preprint, floatfix, superscriptaddress, nofootinbib, longbibliography]{revtex4-2}

\usepackage{graphicx}

\def\vp{\varphi}
\def\ra{\rightarrow}
\def\ds{d^s\!x}
\def\half{\textstyle{\frac{1}{2}}}

\def\ra{\rightarrow}

\def\k{\kappa}

\def\b{\begin{eqnarray*}}  
\def\e{\end{eqnarray*}}    
\def\bn{\begin{eqnarray}}  
\def\en{\end{eqnarray}}   

\begin{document} 
\title{Scaled Affine Quantization of $\vp^4_4$ in the Low Temperature Limit}

\author{Riccardo Fantoni}
\email{riccardo.fantoni@posta.istruzione.it}
\affiliation{Universit\`a di Trieste, Dipartimento di Fisica, \\
strada Costiera 11, 34151 Grignano (Trieste), Italy}
\author{John R. Klauder}
\email{klauder@ufl.edu}
\affiliation{Department of Physics and Department of Mathematics \\
University of Florida,   
Gainesville, FL 32611-8440}

\date{\today}

\begin{abstract}
We prove through Monte Carlo analysis that the covariant euclidean scalar field theory, 
$\vp^r_n$, where $r$ denotes the power of the interaction term and $n = s + 1$ where $s$ 
is the spatial dimension and $1$ adds imaginary time, such that $r = n = 4$ can be 
acceptably quantized using scaled affine quantization and the resulting theory is 
nontrivial and renormalizable even at low temperatures in the highly quantum regime.
\end{abstract}

\maketitle
\section{Introduction}
The classical limit only imposes a constraint on the quantum theory of a given system so 
there is no reason why the classical limit should determine the quantum theory uniquely. 
Accordingly, it is worthwhile to look for alternative quantization recipes, such as 
affine quantization.
We recently showed \cite{Fantoni2020a,Fantoni2020b,Fantoni2021,Fantoni2021b}
that a covariant euclidean scalar field quantization, henceforth denoted $\vp^r_n$, where 
$r$ is the power of the interaction term and $n = s + 1$, where $s$ is the spatial 
dimension and $1$ adds imaginary time, such that $r = 2n/(n - 2)$, e.g., $r = n = 4$, can 
be acceptably quantized using scaled {\it affine quantization} (AQ) 
\cite{Klauder2000,Klauder2020c} and the resulting theory is nontrivial, unlike what 
happens using the usual canonical quantization (CQ) 
\cite{Freedman1982,Aizenman1981,Frohlich1982,Siefert2014}. 
\footnote{In a CQ covariant model the interaction term $g\int \phi(x)^r\, d^nx$ has a power 
$r/n$ per integration. This should be compared with the kinetic term 
$\int [\nabla\phi(x)]^2\, d^nx$ which has a power $2/(n-2)$ per integration. Now, since we 
work in a finite volume region, if $r/n>2/(n-2)$ then the domain where the CQ action is 
finite ${\cal D}_{g>0} \subset {\cal D}_{g=0}$ and the domains change because of reducing $g$ 
back to zero will only retain the smallest version of the domain by continuity, and that will 
not be the theory you started out with so that the CQ model is trivial. Models for which 
$r > 2n/(n - 2)$ have been also recently correctly quantized, as for example $\vp^{12}_3$ 
\cite{Fantoni2020,Fantoni2022}.}
In such studies the temperature was kept constant throughout the whole analysis. It is 
therefore important to study the behavior of the system as we allow temperature to become 
lower and lower thereby approaching the extreme quantum regime. 

The present study will show, through a path integral Monte Carlo (MC) analysis, that as 
the temperature is lowered the renormalized mass is almost unaffected but the 
renormalized coupling constant diminishes. Nonetheless at any given temperature, even in 
the low temperature, strongly quantum, regime, the scaled AQ model appears to  be 
renormalizable showing a non-free behavior in the continuum limit. This success of scaled 
AQ suggests that for the $\vp^4_4$ field theory the more common CQ should be replaced by 
the less known AQ.  

\section{Affine quantization field theory}
For a single scalar field, with spacial degrees of freedom $x=(x_1,x_2,\ldots,x_s)$, 
$\vp(x)$ with canonical momentum $\pi(x)$, the classical affine 
variables are $\k(x)\equiv \pi(x)\,\vp(x)$ and $\vp(x)\neq 0$. The reason we insist that 
$\vp(x)\neq0$ is because if $\vp(x)=0$ then $\k(x)=0$ and $\pi(x)$ can not help.
   
We next introduce the classical Hamiltonian expressed in affine variables. This leads us 
to
\bn \label{eq:affine-H}
{\cal H}(\k,\vp) =\int\{\half[\k(x)^2\,\vp(x)^{-2}+(\nabla\vp(x))^2+m^2\,\vp(x)^2]+g\,\vp(x)^r\}\;\ds, 
\en
where $r$ is a positive, even, integer and $g\geq 0$ is the bare coupling constant such 
that for $g\to 0$ we fall into the free field theory. With these variables we do not let 
$\vp(x)=\infty$ otherwise $\vp(x)^{-2}=0$ which is not fair to $\k(x)$ and, as we already 
observed, we must forbid also $\vp(x)=0$ which would admit $\vp(x)^{-2}=\infty$ giving 
again an undetermined kinetic term. Therefore the AQ bounds $0<|\vp(x)|<\infty$ {\it 
forbid any nonrenormalizability} which is otherwise possible for CQ
\cite{Freedman1982,Aizenman1981,Frohlich1982,Siefert2014}.

The quantum affine operators are the scalar field $\hat{\vp}(x)=\vp(x)$ and the 
{\it dilation} operator 
$\hat{\k}(x)=[\hat{\vp}(x)\hat{\pi}(x)+\hat{\pi}(x)\hat{\vp}(x)]/2$ 
where the momentum operator is $\hat{\pi}(x)=-i\hbar\delta/\delta\vp(x)$. Accordingly for 
the self adjoint kinetic term 
$\hat{\k}(x)\hat{\vp}(x)^{-2}\hat{\k}(x)=\hat{\pi}(x)^2+(3/4)\hbar\delta(0)^{2s}\vp(x)^{-2}$ and one finds for the quantum Hamiltonian operator
\bn \label{eq:HO}
\hat{H}(\hat{\k},\hat{\vp}) =\int\left\{\half[\hat{\pi}(x)^2+(\nabla\vp(x))^2+m^2\,\vp(x)^2]+g\,\vp(x)^r +{\textstyle\frac{3}{8}}\hbar^2\frac{\delta(0)^{2s}}{\vp(x)^2}\right\}\;\ds.
\en

The affine action is found adding time, $x_0=ct$, where $c$ is the speed of light 
constant and $t$ is imaginary time, so that ${\cal S}=\int_0^\beta H\,dx_0$, with $H$ the 
semi-classical Hamiltonian corresponding to the one of Eq. (\ref{eq:HO}), will then read
\bn \label{eq:action}
{\cal S}[\vp]=\int_0^\beta dx_0\,\int_{L^s}d^sx\,\left\{\frac{1}{2}\left[\sum_{\mu=0}^s\left(\frac{\partial\vp(x)}{\partial x_\mu}\right)^2+m^2\,\vp(x)^2\right]+g\,\vp(x)^r+\frac{3}{8}\hbar\frac{\delta(0)^{2s}}{\vp(x)^2}\right\}, 
\en
where with an abuse of notation we here use $x$ for $(x_0,x_1,x_2,\ldots,x_s)$ and 
$\beta=1/k_BT$, with $k_B$ the Boltzmann's constant, is the inverse temperature.
At low temperatures the quantum effects become more relevant and this is the regime we 
are interested in this work. 

The vacuum expectation value of an observable ${\cal O}[\vp]$ will then be given by the 
following expression
\bn \label{eq:EV}
\langle{\cal O}\rangle=\frac{\int{\cal O}[\vp]\exp(-{\cal S}[\vp])\;{\cal D}\vp(x)}{\int\exp(-{\cal S}[\vp])\;{\cal D}\vp(x)},
\en
where the functional integrals will be calculated on a lattice using the path integral 
Monte Carlo method as explained further on.

\section{Lattice formulation of the field theory}
The theory considers a real scalar field $\vp$ taking the value $\vp(x)$ on each site of 
a periodic $n$-dimensional lattice, with $n=s+1$ space-time dimensions, of lattice 
spacing $a$, the ultraviolet cutoff, and spacial periodicity $L=Na$ and temporal 
periodicity $\beta=N_0a$. The field path is a closed loop on an $n$-dimensional surface 
of an $(n+1)$-dimensional $\beta$-cylinder.
We used a lattice formulation of the AQ field theory of Eq. (\ref{eq:action}) (also 
studied in Eq. (8) of \cite{Fantoni2020a}) using the scaling $\vp\ra a^{-s/2}\vp$ and 
$g\ra a^{s(r-2)/2} g$ which is necessary
\footnote{Note that from a physical point of view one never has to worry about the 
mathematical divergence since the lattice spacing will necessarily have a lower bound. For 
example at an atomic level one will have $a\gtrsim 1$\AA. In other words the continuum limit 
will never be a mathematical one.}
to eliminate the Dirac delta factor 
$\delta(0)=a^{-1}$ divergent in the continuum limit $a\to 0$. The affine action for the 
field (in the {\it primitive approximation} \cite{Ceperley1995}) has then the following valid 
discretization
\bn \label{eq:scaled-affine-action}
S[\vp]/a=\half\left\{\sum_{x,\mu}a^{-2}[\vp(x)-\vp(x+e_\mu)]^2 
+m^2\sum_{x}\vp(x)^2\right\}+\sum_{x}g\,\vp(x)^r+{\textstyle\frac{3}{8}\sum_{x}}{\displaystyle\frac{\hbar^2}{\vp(x)^2}},
\en
where $e_\mu$ is a vector of length $a$ in the $+\mu$ direction with 
$\mu=0,1,2,\ldots,s$. We will have ${\cal S}\approx S$.

In this work we are interested in reaching the continuum limit by taking $Na$ fixed 
and letting $N\to\infty$ at fixed volume $L^s$. The absolute temperature $T=1/k_B\beta$ 
is allowed to vary so that the number of discretization points for the imaginary time 
interval $[0,\beta]$ will be $N_0=\beta/a$. We are here interested in the $N_0\gg N$ (or 
$\beta\gg L$) regime. 

\subsection{Monte Carlo results}
We performed a path integral MC \cite{Metropolis,Kalos-Whitlock,Ceperley1995,Fantoni12d} 
calculation for the AQ field theory described by the action of Eq. 
(\ref{eq:scaled-affine-action}). We calculated the renormalized coupling 
constant $g_R$ and mass $m_R$ defined in Eqs. (11) and (13) of \cite{Fantoni2020a} 
respectively, measuring them in the path integral MC through vacuum expectation values 
like in Eq. (\ref{eq:EV}). In particular 
\bn
m_R^2=\frac{p_0^2\langle|\tilde{\vp}(p_0)|^2\rangle}{\langle\tilde{\vp}(0)^2\rangle-\langle|\tilde{\vp}(p_0)|^2\rangle},
\en 
and at zero momentum
\bn
g_R=\frac{3\langle\tilde{\vp}(0)^2\rangle^2-\langle\tilde{\vp}(0)^4\rangle}{\langle\tilde{\vp}(0)^2\rangle^2},
\en
where $\tilde{\vp}(p)=\int d^nx\;e^{ip\cdot x}\vp(x)$ is the Fourier transform of the field 
and we choose the 4-momentum $p_0$ with one spacial component equal to $2\pi/Na$ and all 
other components equal to zero.

In our previous studies \cite{Fantoni2020a,Fantoni2021b} we set $L=\beta=1$. Here we will 
consider $L=1$ and $\beta\gg L$ instead. As usual we will impose periodic boundary 
conditions both in space and in imaginary time. We will use natural units 
$c=\hbar=k_B=1$ throughout the whole analysis.

Following Freedman et al. \cite{Freedman1982}, we fix (within 10\%) the renormalized mass 
$m_R\approx 3$, tuning appropriately the bare mass $m$ by trial and error, and we measure 
the renormalized coupling constant $g_R$ at various values of 
the bare coupling $g$. We found that the renormalized mass is almost independent on 
$\beta$. So we chose the same values of $m$ for all the temperatures studied. But the 
renormalized coupling $g_R$ diminishes as $\beta$ and/or $m$ increase. It is then 
convenient to define a second renormalized coupling constant which is less 
dependent on $\beta, L,$ and $m$. Following Freedman et al. \cite{Freedman1982} we set 
$G_R=g_Rm_R^nL^s\beta$. 

\begin{figure}[htbp]
\begin{center}
\includegraphics[width=5.5cm]{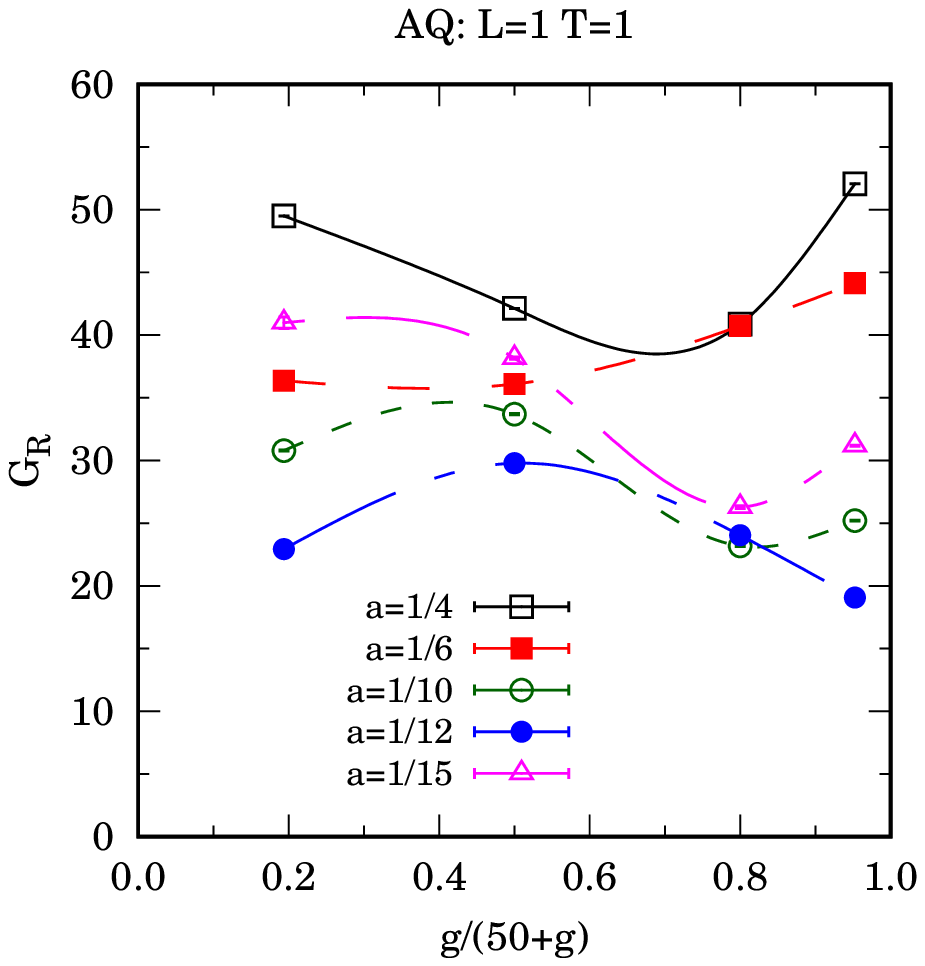}\includegraphics[width=5.5cm]{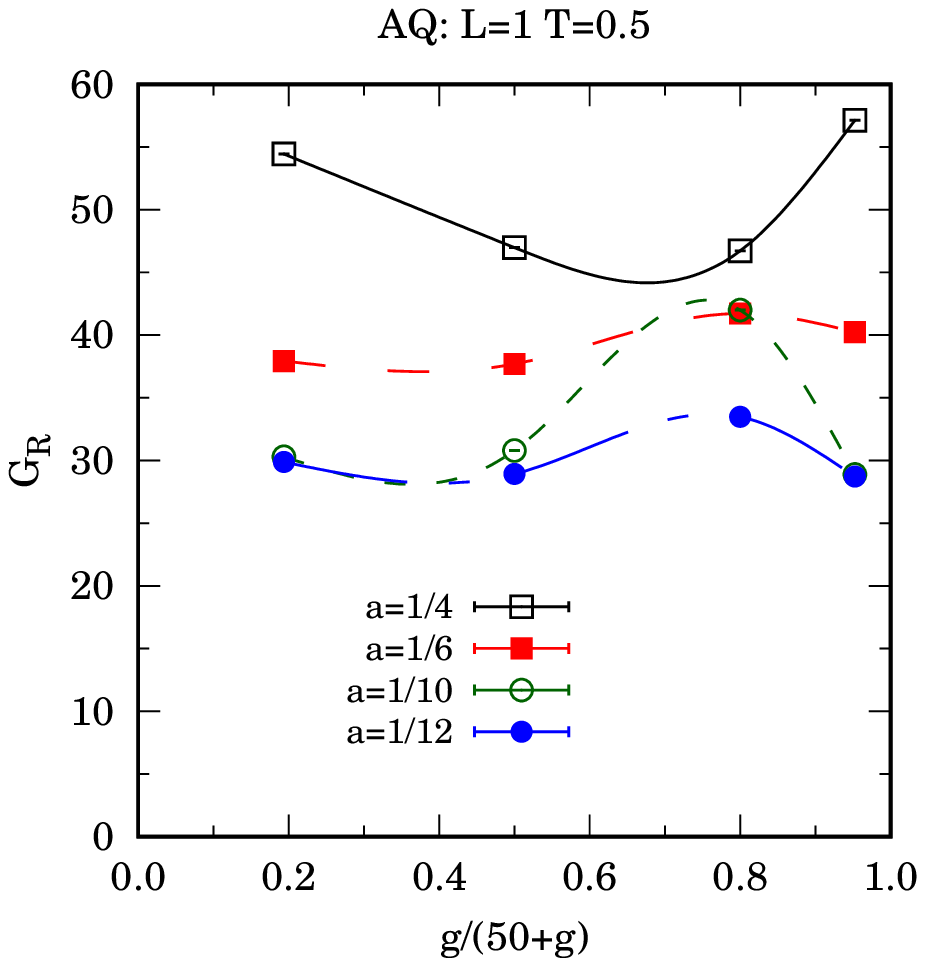}\includegraphics[width=5.5cm]{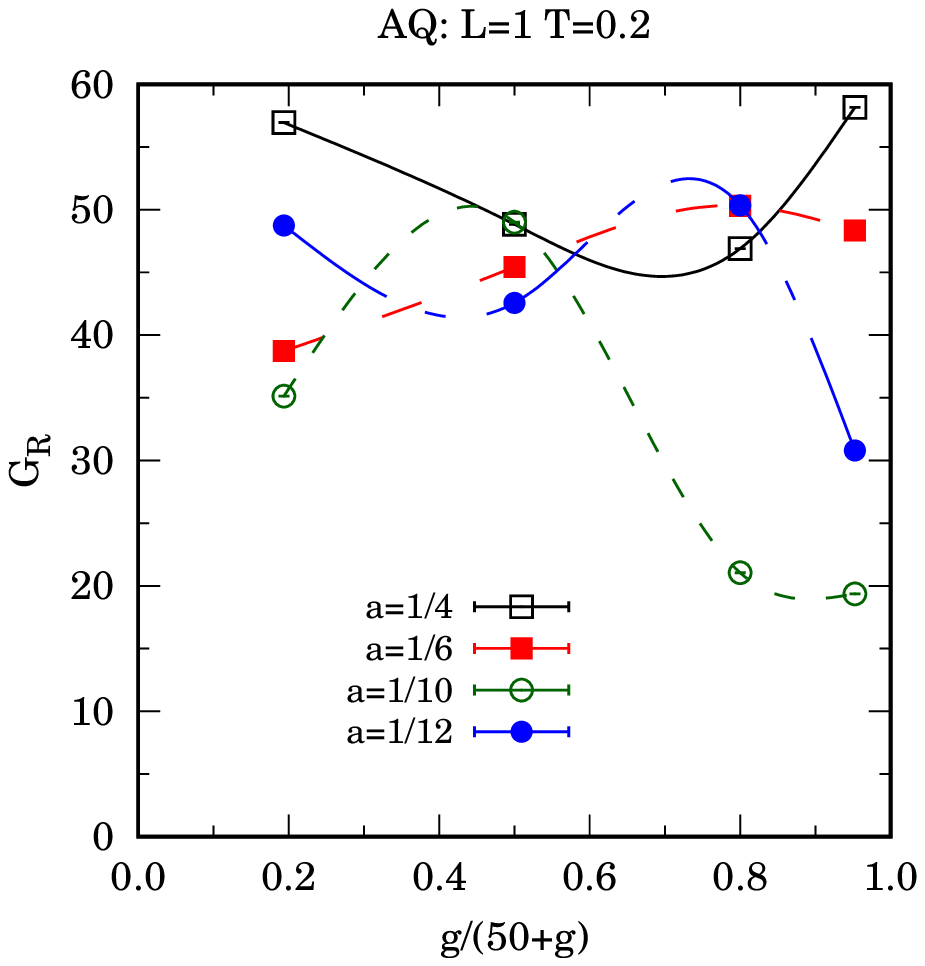}
\end{center}
\caption{(color online) The left panel is for AQ with $L=1$ and $T=1$, The central panel 
is for AQ with $L=1$ and $T=0.5$, and the right panel is for AQ with $L=1$ and $T=0.2$. 
We show the renormalized coupling constant $G_R$, defined in the text, as a function of 
$g/(50+g)$ for decreasing values of the lattice spacing $a$. The renormalized mass was 
kept fixed to $m_R\approx 3$ (within 10\%) in all cases. The statistical errors in the 
Monte Carlo were in all cases smaller than the symbols used. The main source of 
uncertainty is nonetheless the indirect one stemming from the unavoidable difficulty of 
keeping the renormalized mass constant throughout all cases. The lines connecting the 
points are just a guide for the eye.}
\label{fig:a4-4}
\end{figure}

We chose two low temperatures (the case $T=1$ had already been studied in Ref. 
\cite{Fantoni2021b}), namely an intermediate one $T=0.5$ and an extreme one 
$T=0.2$. In each case we study the continuum limit by choosing decreasing values of $a$, 
namely $a=1/4,1/6,1/10,1/12$ and $1/15$ corresponding respectively to 
$N_0=1/Ta=8,12,20,24,30$ for $T=0.5$ and to $N_0=20,30,50,60,75$ for $T=0.2$. In each run we 
used $3\times 10^7$ MC steps, where one step consists in $N^sN_0$ Metropolis 
\cite{Metropolis} configuration moves of each field component, reaching equilibrium after 
10\% of the largest $a$ run to 50\% of the smallest $a$ run. In our simulations we used block 
averages and estimated the statistical errors using the jakknife method (described in Section 
3.6 of \cite{Janke2002}) to take into account of the correlation time. It took roughly 25 
days of computer time for the $T=0.2, a=1/12$ run to complete. In Fig. \ref{fig:a4-4} we show 
the numerical results.

From the figure we can see how at all temperatures and all bare coupling constants $G_R$ 
tends to stay far from zero as we approach the continuum limit $a\to 0$. Moreover, 
with respect to the case $T=1$, already studied in Ref. \cite{Fantoni2021b}, where the 
value for $G_R$ tends to revert its trend to decrease for a decrease of the lattice spacing 
only for an ultraviolet cutoff as small as $a=1/15$, now we find that at $T=0.5$ this 
inversion happens already for $a=1/10$ at least at intermediate bare coupling and at $T=0.2$ 
already for $a=1/6$. This had to be expected on general grounds because it is impossible to 
distinguish time from the other spacial components just by looking at the action expression 
(\ref{eq:scaled-affine-action}) 
and the $T=1, a=1/15$ case has a total of $15^4=50625$ lattice points which is very close 
to the total lattice points of the case $T=0.2, a=1/10$ which are $10^350=50000$. We are 
just choosing an hyperrectangle instead of an hypercube periodic lattice. Nonetheless 
there is a strong indication that our scaled AQ model is indeed non-free in the continuum 
thus resulting renormalizable, unlike the corresponding CQ model \footnote{For a 
comparison with the corresponding scaled CQ results see Ref. \citep{Fantoni2022b} and for 
the unscaled CQ ones see Ref. \cite{Fantoni2020a}}. And the more so at lower 
temperatures. We can therefore infer that the same should continue to hold also in the 
$T\to 0$, ground state, limit. 

\section{Conclusions}
In conclusion we studied the renormalizability property of one real scalar covariant 
euclidean field quantized through scaled affine quantization (AQ) with the path integral 
Monte Carlo method on a lattice permeating the whole spacetime. We therefore used 
periodic spacial boundary conditions at finite unit volume to simulate an infinite volume 
system and in measuring the renormalized mass and coupling constant of the model we also 
enforced periodic temporal boundary conditions which are necessary in order to determine 
the required vacuum expectation values. The periodicity on the imaginary time, i.e. 
the inverse temperature $\beta=1/T$, was chosen at increasing values equal to $1,2,5$.  
Keeping fixed the renormalized mass, our numerical results for the renormalized 
coupling constant showed how this has a non monotonically decreasing behavior with respect 
to a decreasing lattice spacing. This remains true even at low temperature thus proving the 
renormalizability of the model even when the temperature is lowered in the extreme 
quantum regime. We therefore suspect that the non triviality still holds for the ground 
state.

On general grounds we should accept affine quantization as a way to remove infinities, 
which are mathematical but not physical, from the field theory. In fact just by looking 
at the kinetic term in Eq. (\ref{eq:affine-H}) we can say that if $\vp$ is allowed to 
become infinity (or zero) then $\k$ cannot help. If $\k$ becomes infinite then $\vp$ 
cannot help.  $\k=0$ is allowed so that $\pi=0$. When $\pi$ and $\vp$ were alone, as in 
the canonical quantization picture, they could allow mathematical infinities. In a 
physical (or Monte Carlo) measure of an observable there is no space for mathematical 
infinities.

For the Higgs sector of the Standard Model, the low energy properties are very specific 
and, so far, observation confirms that they are well described by canonical $\vp^4$. It 
is certainly true that canonical quantization (CQ) of $\vp^4$ does not reach down to 
distances of the order of the Planck length -- in that realm, anyway, gravity cannot be 
dealt with classically -- so affine quantization (AQ) may be used to solve this problem.

\bibliography{four-four}

\begin{thebibliography}{18}%
\makeatletter
\providecommand \@ifxundefined [1]{%
 \@ifx{#1\undefined}
}%
\providecommand \@ifnum [1]{%
 \ifnum #1\expandafter \@firstoftwo
 \else \expandafter \@secondoftwo
 \fi
}%
\providecommand \@ifx [1]{%
 \ifx #1\expandafter \@firstoftwo
 \else \expandafter \@secondoftwo
 \fi
}%
\providecommand \natexlab [1]{#1}%
\providecommand \enquote  [1]{``#1''}%
\providecommand \bibnamefont  [1]{#1}%
\providecommand \bibfnamefont [1]{#1}%
\providecommand \citenamefont [1]{#1}%
\providecommand \href@noop [0]{\@secondoftwo}%
\providecommand \href [0]{\begingroup \@sanitize@url \@href}%
\providecommand \@href[1]{\@@startlink{#1}\@@href}%
\providecommand \@@href[1]{\endgroup#1\@@endlink}%
\providecommand \@sanitize@url [0]{\catcode `\\12\catcode `\$12\catcode
  `\&12\catcode `\#12\catcode `\^12\catcode `\_12\catcode `\%12\relax}%
\providecommand \@@startlink[1]{}%
\providecommand \@@endlink[0]{}%
\providecommand \url  [0]{\begingroup\@sanitize@url \@url }%
\providecommand \@url [1]{\endgroup\@href {#1}{\urlprefix }}%
\providecommand \urlprefix  [0]{URL }%
\providecommand \Eprint [0]{\href }%
\providecommand \doibase [0]{https://doi.org/}%
\providecommand \selectlanguage [0]{\@gobble}%
\providecommand \bibinfo  [0]{\@secondoftwo}%
\providecommand \bibfield  [0]{\@secondoftwo}%
\providecommand \translation [1]{[#1]}%
\providecommand \BibitemOpen [0]{}%
\providecommand \bibitemStop [0]{}%
\providecommand \bibitemNoStop [0]{.\EOS\space}%
\providecommand \EOS [0]{\spacefactor3000\relax}%
\providecommand \BibitemShut  [1]{\csname bibitem#1\endcsname}%
\let\auto@bib@innerbib\@empty
\bibitem [{\citenamefont {Fantoni}\ and\ \citenamefont
  {Klauder}(2021{\natexlab{a}})}]{Fantoni2020a}%
  \BibitemOpen
  \bibfield  {author} {\bibinfo {author} {\bibfnamefont {R.}~\bibnamefont
  {Fantoni}}\ and\ \bibinfo {author} {\bibfnamefont {J.~R.}\ \bibnamefont
  {Klauder}},\ }\bibfield  {title} {\bibinfo {title} {{Affine Quantization of
  $(\varphi^{4})_4$ Succeeds While Canonical Quantization Fails}},\ }\href@noop
  {} {\bibfield  {journal} {\bibinfo  {journal} {Phys. Rev. D}\ }\textbf
  {\bibinfo {volume} {103}},\ \bibinfo {pages} {076013} (\bibinfo {year}
  {2021}{\natexlab{a}})}\BibitemShut {NoStop}%
\bibitem [{\citenamefont {Fantoni}\ and\ \citenamefont
  {Klauder}(2021{\natexlab{b}})}]{Fantoni2020b}%
  \BibitemOpen
  \bibfield  {author} {\bibinfo {author} {\bibfnamefont {R.}~\bibnamefont
  {Fantoni}}\ and\ \bibinfo {author} {\bibfnamefont {J.~R.}\ \bibnamefont
  {Klauder}},\ }\bibfield  {title} {\bibinfo {title} {{Monte Carlo evaluation
  of the continuum limit of the two-point function of the Euclidean free real
  scalar field subject to affine quantization}},\ }\href@noop {} {\bibfield
  {journal} {\bibinfo  {journal} {J. Stat. Phys.}\ }\textbf {\bibinfo {volume}
  {184}},\ \bibinfo {pages} {28} (\bibinfo {year}
  {2021}{\natexlab{b}})}\BibitemShut {NoStop}%
\bibitem [{\citenamefont {Fantoni}\ and\ \citenamefont
  {Klauder}(2021{\natexlab{c}})}]{Fantoni2021}%
  \BibitemOpen
  \bibfield  {author} {\bibinfo {author} {\bibfnamefont {R.}~\bibnamefont
  {Fantoni}}\ and\ \bibinfo {author} {\bibfnamefont {J.~R.}\ \bibnamefont
  {Klauder}},\ }\bibfield  {title} {\bibinfo {title} {{Monte Carlo evaluation
  of the continuum limit of the two-point function of two Euclidean Higgs real
  scalar field subject to affine quantization}},\ }\href@noop {} {\bibfield
  {journal} {\bibinfo  {journal} {Phys. Rev. D}\ }\textbf {\bibinfo {volume}
  {104}},\ \bibinfo {pages} {054514} (\bibinfo {year}
  {2021}{\natexlab{c}})}\BibitemShut {NoStop}%
\bibitem [{\citenamefont {Fantoni}\ and\ \citenamefont
  {Klauder}(2022{\natexlab{a}})}]{Fantoni2021b}%
  \BibitemOpen
  \bibfield  {author} {\bibinfo {author} {\bibfnamefont {R.}~\bibnamefont
  {Fantoni}}\ and\ \bibinfo {author} {\bibfnamefont {J.~R.}\ \bibnamefont
  {Klauder}},\ }\bibfield  {title} {\bibinfo {title} {{Eliminating
  Nonrenormalizability Helps Prove Scaled Affine Quantization of $\varphi^4_4$
  is Nontrivial}},\ }\href@noop {} {\bibfield  {journal} {\bibinfo  {journal}
  {Int. J. Mod. Phys. A}\ }\textbf {\bibinfo {volume} {37}},\ \bibinfo {pages}
  {2250029} (\bibinfo {year} {2022}{\natexlab{a}})}\BibitemShut {NoStop}%
\bibitem [{\citenamefont {Klauder}(2000)}]{Klauder2000}%
  \BibitemOpen
  \bibfield  {author} {\bibinfo {author} {\bibfnamefont {J.~R.}\ \bibnamefont
  {Klauder}},\ }\href@noop {} {\emph {\bibinfo {title} {{Beyond Conventional
  Quantization}}}}\ (\bibinfo  {publisher} {Cambridge University Press},\
  \bibinfo {year} {2000})\ \bibinfo {note} {chap. 5}\BibitemShut {NoStop}%
\bibitem [{\citenamefont {Klauder}(2020)}]{Klauder2020c}%
  \BibitemOpen
  \bibfield  {author} {\bibinfo {author} {\bibfnamefont {J.~R.}\ \bibnamefont
  {Klauder}},\ }\bibfield  {title} {\bibinfo {title} {{The Benefits of Affine
  Quantization}},\ }\href@noop {} {\bibfield  {journal} {\bibinfo  {journal}
  {Journal of High Energy Physics, Gravitation and Cosmology}\ }\textbf
  {\bibinfo {volume} {6}},\ \bibinfo {pages} {175} (\bibinfo {year}
  {2020})}\BibitemShut {NoStop}%
\bibitem [{\citenamefont {Freedman}\ \emph {et~al.}(1982)\citenamefont
  {Freedman}, \citenamefont {Smolensky},\ and\ \citenamefont
  {Weingarten}}]{Freedman1982}%
  \BibitemOpen
  \bibfield  {author} {\bibinfo {author} {\bibfnamefont {B.}~\bibnamefont
  {Freedman}}, \bibinfo {author} {\bibfnamefont {P.}~\bibnamefont
  {Smolensky}},\ and\ \bibinfo {author} {\bibfnamefont {D.}~\bibnamefont
  {Weingarten}},\ }\bibfield  {title} {\bibinfo {title} {{Monte Carlo
  Evaluation of the Continuum Limit of $\phi_4^4$ and $\phi_3^4$}},\
  }\href@noop {} {\bibfield  {journal} {\bibinfo  {journal} {Physics Letters}\
  }\textbf {\bibinfo {volume} {113B}},\ \bibinfo {pages} {481} (\bibinfo {year}
  {1982})}\BibitemShut {NoStop}%
\bibitem [{\citenamefont {Aizenman}(1981)}]{Aizenman1981}%
  \BibitemOpen
  \bibfield  {author} {\bibinfo {author} {\bibfnamefont {M.}~\bibnamefont
  {Aizenman}},\ }\bibfield  {title} {\bibinfo {title} {{Proof of the Triviality
  of $\phi^4_d$ Field Theory and Some Mean-Field Features of Ising Models for
  $d > 4$}},\ }\href@noop {} {\bibfield  {journal} {\bibinfo  {journal} {Phys.
  Rev. Lett.}\ }\textbf {\bibinfo {volume} {47}},\ \bibinfo {pages} {886(E)}
  (\bibinfo {year} {1981})}\BibitemShut {NoStop}%
\bibitem [{\citenamefont {Fr\"ohlich}(1982)}]{Frohlich1982}%
  \BibitemOpen
  \bibfield  {author} {\bibinfo {author} {\bibfnamefont {J.}~\bibnamefont
  {Fr\"ohlich}},\ }\bibfield  {title} {\bibinfo {title} {{On the Triviality of
  $\lambda\phi_d^4$ Theories and the Approach to the Critical Point in $d\geq
  4$ Dimensions}},\ }\href@noop {} {\bibfield  {journal} {\bibinfo  {journal}
  {Nuclear Physics B}\ }\textbf {\bibinfo {volume} {200}},\ \bibinfo {pages}
  {281} (\bibinfo {year} {1982})}\BibitemShut {NoStop}%
\bibitem [{\citenamefont {Siefert}\ and\ \citenamefont
  {Wolff}(2014)}]{Siefert2014}%
  \BibitemOpen
  \bibfield  {author} {\bibinfo {author} {\bibfnamefont {J.}~\bibnamefont
  {Siefert}}\ and\ \bibinfo {author} {\bibfnamefont {U.}~\bibnamefont
  {Wolff}},\ }\bibfield  {title} {\bibinfo {title} {{Triviality of $\varphi^4$
  theory in a finite volume scheme adapted to the broken phase}},\ }\href@noop
  {} {\bibfield  {journal} {\bibinfo  {journal} {Physics Letters B}\ }\textbf
  {\bibinfo {volume} {733}},\ \bibinfo {pages} {11} (\bibinfo {year}
  {2014})}\BibitemShut {NoStop}%
\bibitem [{\citenamefont {Fantoni}(2021)}]{Fantoni2020}%
  \BibitemOpen
  \bibfield  {author} {\bibinfo {author} {\bibfnamefont {R.}~\bibnamefont
  {Fantoni}},\ }\bibfield  {title} {\bibinfo {title} {{Monte Carlo Evaluation
  of the Continuum Limit of $(\phi^{12})_3$}},\ }\href@noop {} {\bibfield
  {journal} {\bibinfo  {journal} {J. Stat. Mech.}\ ,\ \bibinfo {pages}
  {083102}} (\bibinfo {year} {2021})}\BibitemShut {NoStop}%
\bibitem [{\citenamefont {Fantoni}(2022)}]{Fantoni2022}%
  \BibitemOpen
  \bibfield  {author} {\bibinfo {author} {\bibfnamefont {R.}~\bibnamefont
  {Fantoni}},\ }\bibfield  {title} {\bibinfo {title} {{Scaled Affine
  Quantization of $\varphi^{12}_3$ is Nontrivial}},\ }\href@noop {} {\bibfield
  {journal} {\bibinfo  {journal} {Eur. Phys. J. C (submitted)}\ } (\bibinfo
  {year} {2022})}\BibitemShut {NoStop}%
\bibitem [{\citenamefont {{D. M. Ceperley}}(1995)}]{Ceperley1995}%
  \BibitemOpen
  \bibfield  {author} {\bibinfo {author} {\bibnamefont {{D. M. Ceperley}}},\
  }\href@noop {} {\bibfield  {journal} {\bibinfo  {journal} {Rev. Mod. Phys.}\
  }\textbf {\bibinfo {volume} {67}},\ \bibinfo {pages} {279} (\bibinfo {year}
  {1995})}\BibitemShut {NoStop}%
\bibitem [{\citenamefont {Metropolis}\ \emph {et~al.}(1953)\citenamefont
  {Metropolis}, \citenamefont {Rosenbluth}, \citenamefont {Rosenbluth},
  \citenamefont {Teller},\ and\ \citenamefont {Teller}}]{Metropolis}%
  \BibitemOpen
  \bibfield  {author} {\bibinfo {author} {\bibfnamefont {N.}~\bibnamefont
  {Metropolis}}, \bibinfo {author} {\bibfnamefont {A.~W.}\ \bibnamefont
  {Rosenbluth}}, \bibinfo {author} {\bibfnamefont {M.~N.}\ \bibnamefont
  {Rosenbluth}}, \bibinfo {author} {\bibfnamefont {A.~M.}\ \bibnamefont
  {Teller}},\ and\ \bibinfo {author} {\bibfnamefont {E.}~\bibnamefont
  {Teller}},\ }\bibfield  {title} {\bibinfo {title} {{Equation of State
  Calculations by Fast Computing Machines}},\ }\href@noop {} {\bibfield
  {journal} {\bibinfo  {journal} {J. Chem. Phys.}\ }\textbf {\bibinfo {volume}
  {1087}},\ \bibinfo {pages} {21} (\bibinfo {year} {1953})}\BibitemShut
  {NoStop}%
\bibitem [{\citenamefont {Kalos}\ and\ \citenamefont
  {Whitlock}(2008)}]{Kalos-Whitlock}%
  \BibitemOpen
  \bibfield  {author} {\bibinfo {author} {\bibfnamefont {M.~H.}\ \bibnamefont
  {Kalos}}\ and\ \bibinfo {author} {\bibfnamefont {P.~A.}\ \bibnamefont
  {Whitlock}},\ }\href@noop {} {\emph {\bibinfo {title} {{Monte Carlo
  Methods}}}}\ (\bibinfo  {publisher} {Wiley-Vch Verlag GmbH \& Co.},\ \bibinfo
  {address} {Germany},\ \bibinfo {year} {2008})\BibitemShut {NoStop}%
\bibitem [{\citenamefont {Fantoni}(2012)}]{Fantoni12d}%
  \BibitemOpen
  \bibfield  {author} {\bibinfo {author} {\bibfnamefont {R.}~\bibnamefont
  {Fantoni}},\ }\bibfield  {title} {\bibinfo {title} {{Localization of acoustic
  polarons at low temperatures: A path integral Monte Carlo approach}},\ }\href
  {https://doi.org/10.1103/PhysRevB.86.144304} {\bibfield  {journal} {\bibinfo
  {journal} {Phys. Rev. B}\ }\textbf {\bibinfo {volume} {86}},\ \bibinfo
  {pages} {144304} (\bibinfo {year} {2012})}\BibitemShut {NoStop}%
\bibitem [{\citenamefont {Janke}(2002)}]{Janke2002}%
  \BibitemOpen
  \bibfield  {author} {\bibinfo {author} {\bibfnamefont {W.}~\bibnamefont
  {Janke}},\ }\bibfield  {title} {\bibinfo {title} {{Statistical Analysis of
  Simulations: Data Correlations and Error Estimation}},\ }in\ \href@noop {}
  {\emph {\bibinfo {booktitle} {{Quantum Simulations of Complex Many-Body
  Systems: From Theory to Algorithms}}}},\ \bibinfo {series} {NIC Series},
  Vol.~\bibinfo {volume} {10},\ \bibinfo {editor} {edited by\ \bibinfo {editor}
  {\bibfnamefont {J.}~\bibnamefont {Grotendorst}}, \bibinfo {editor}
  {\bibfnamefont {D.}~\bibnamefont {Marx}},\ and\ \bibinfo {editor}
  {\bibfnamefont {A.}~\bibnamefont {Muramatsu}}}\ (\bibinfo  {publisher} {John
  von Neumann Institute for Computing},\ \bibinfo {address} {J\"ulich},\
  \bibinfo {year} {2002})\ pp.\ \bibinfo {pages} {423--445},\ \bibinfo {note}
  {{ISBN 3-00-009057-6}}\BibitemShut {NoStop}%
\bibitem [{\citenamefont {Fantoni}\ and\ \citenamefont
  {Klauder}(2022{\natexlab{b}})}]{Fantoni2022b}%
  \BibitemOpen
  \bibfield  {author} {\bibinfo {author} {\bibfnamefont {R.}~\bibnamefont
  {Fantoni}}\ and\ \bibinfo {author} {\bibfnamefont {J.~R.}\ \bibnamefont
  {Klauder}},\ }\bibfield  {title} {\bibinfo {title} {{Kinetic Factors in
  Affine Quantization and Their Role in Field Theory Monte Carlo}},\
  }\href@noop {} {\bibfield  {journal} {\bibinfo  {journal} {Int. J. Mod. Phys.
  A}\ }\textbf {\bibinfo {volume} {37}},\ \bibinfo {pages} {2250094} (\bibinfo
  {year} {2022}{\natexlab{b}})}\BibitemShut {NoStop}%
\end{thebibliography}%


\end{document}